\documentclass[aps,prl,showpacs,reprint,superscriptaddress,twocolumn]{revtex4-1}% Physical Review Letters

\usepackage{txfonts} % for lambda bar

\usepackage{graphicx}% Include figure files
\usepackage{dcolumn}% Align table columns on decimal point
\usepackage{bm}% bold mathmwe
\usepackage{color}% This is just used temporarily, for notes to myself
\usepackage{subfigure}% Allow for grouped figures.
\usepackage{multirow}% Support for tables
\usepackage{amssymb}
\usepackage{hyperref}
\usepackage{eucal}
\usepackage{gensymb}

\newcommand{\iso}[2]{$^{#1}$#2}

\newcommand{\ucb}{Department of Nuclear Engineering, University of California, Berkeley, California 94720, USA}
\newcommand{\llnl}{Lawrence Livermore National Laboratory, Livermore, California 94550, USA}
\newcommand{\psu}{Department of Mechanical and Nuclear Engineering, The Pennsylvania State University, University Park, Pennsylvania 16802, USA}

\begin{document}
\title{First measurement of the ionization yield of nuclear recoils in liquid argon}

\author{T.H.~Joshi} \email {thjoshi@berkeley.edu} \affiliation{\ucb} \affiliation{\llnl}
\author{S.~Sangiorgio} \affiliation{\llnl}
\author{A.~Bernstein} \affiliation{\llnl}
\author{M.~Foxe} \altaffiliation{present address: Pacific Northwest National Laboratory} \affiliation{\llnl} \affiliation{\psu} 
\author{C.~Hagmann} \affiliation{\llnl}
\author{I.~Jovanovic} \affiliation{\psu}
\author{K.~Kazkaz} \affiliation{\llnl}
\author{V.~Mozin} \affiliation{\llnl}
\author{E.B.~Norman} \affiliation{\ucb} \affiliation{\llnl}
\author{S.V.~Pereverzev} \affiliation{\llnl}
\author{F.~Rebassoo} \affiliation {\llnl}
\author{P.~Sorensen} \affiliation{\llnl}

\date{25 April 2014}

\begin{abstract}
This Letter details a measurement of the ionization yield ($Q_y$) of 6.7 keV \iso{40}{Ar} atoms stopping in a liquid argon detector. The $Q_y$ of 3.6--6.3 detected $e^{-}/\mbox{keV}$, for applied electric fields in the range 240--2130 V/cm, is encouraging for the use of this detector medium to search for the signals from hypothetical dark matter particle interactions and from coherent elastic neutrino-nucleus scattering. A significant dependence of $Q_y$ on the applied electric field is observed and explained in the context of ion recombination.
\end{abstract}
\maketitle

% \textbf{Introduction~} 
Liquid-phase argon has long been used as a target medium for particle detection via scintillation and charge collection. 
Recently there has been considerable interest in direct detection of both hypothetical dark matter particles \cite{Gaitskell} and coherent elastic neutrino-nucleus scattering (CENNS) \cite{Freedman,Drukier}. 
These as-yet unobserved neutral particle interactions are expected to result in a recoiling argon atom $\mathcal{O}$(keV), generally referred to in the literature as a nuclear recoil. 
This prompts the question of the available signal produced by such recoils in a liquid argon detector.  This quantity must be directly measured due to the difference in signals from nuclear recoils as opposed to electron recoils (e.g. Compton electrons and $\beta$-particles).
In this Letter we report the first measurement of the ionization yield ($Q_{y}$) (detected electrons per unit energy) resulting from nuclear recoils in liquid argon, measured at 6.7 keV.  
This is also the lowest-energy measurement of nuclear recoils in liquid argon.

These results are of interest not only for particle detection, but for theoretical studies of condensed media as well. 
Models of the production of ions and excited atoms from low-energy recoils in liquid argon exist, but are not fully understood in the few-keV energy range \cite{Chepel}. 
To study the influence of the electric field on recombination, and thus $Q_y$, data were obtained at applied electric field values of 240, 640, 1600, 2130~V/cm.

The scintillation efficiency of nuclear recoils in liquid argon has been measured from 10--250 keV at zero electric drift field using the kinematically constrained scatter of 2.8 MeV neutrons \cite{Gastler:2010sc} and from 11--50 keV at electric drift fields from 0--1000 V/cm using the kinematically constrained scatter of 0.60 and 1.17 MeV neutrons \cite{Alexander}.  
No measurements of nuclear recoils in liquid argon exist below 10 keV.  

Liquid argon dual-phase detectors have been shown to be sensitive to single electrons generated in the bulk \cite{Sangiorgio201369}.
This enhances the detection capability of the ionization channel over the scintillation channel at very low energies.  
A low-energy threshold and calibration are critical in both dark matter searches  and CENNS discovery. 
Both interactions exhibit a recoil energy spectrum that rises rapidly with decreasing energy \cite{Chepel,Hagmann,Akimov}. 
Our results suggest that dark matter searches using only the ionization channel in liquid argon (as has been done in liquid xenon \cite{Angle:2011th}) could probe an interesting new parameter space.  
The observation and modeling of electric drift-field dependence presented in this Letter, and also recently reported in the scintillation channel \cite{Alexander}, lay the foundation for a comprehensive understanding of ion recombination in liquid argon and suggests the need for optimization of drift fields in future liquid argon-based experiments.  

\textit{Experimental Details}.--
Our measurement employed a beam of neutrons to create nuclear recoils in liquid argon.
The neutron spectrum was peaked at 24 and 70 keV. Contributions from the quasimonoenergetic 70 keV (12\% FWHM) neutrons were selected during background subtraction.  
The design and deployment of the neutron beam is described in detail in Ref. \cite{nsource}. 
Our detector, a small dual-phase argon proportional scintillation counter, is described in Ref. \cite{Sangiorgio201369}. 
Small modifications to the detector since that work include the removal of the $^{55}$Fe source and holder, and the replacement of one of the electrode grids. 
The response to \iso{37}{Ar} calibrations has been verified to be consistent with the previous results. 
The active region of liquid argon has a 2.5-cm radius and a 3.7-cm height. 

\begin{figure*}[]
\includegraphics[angle=0,width=0.98\textwidth]{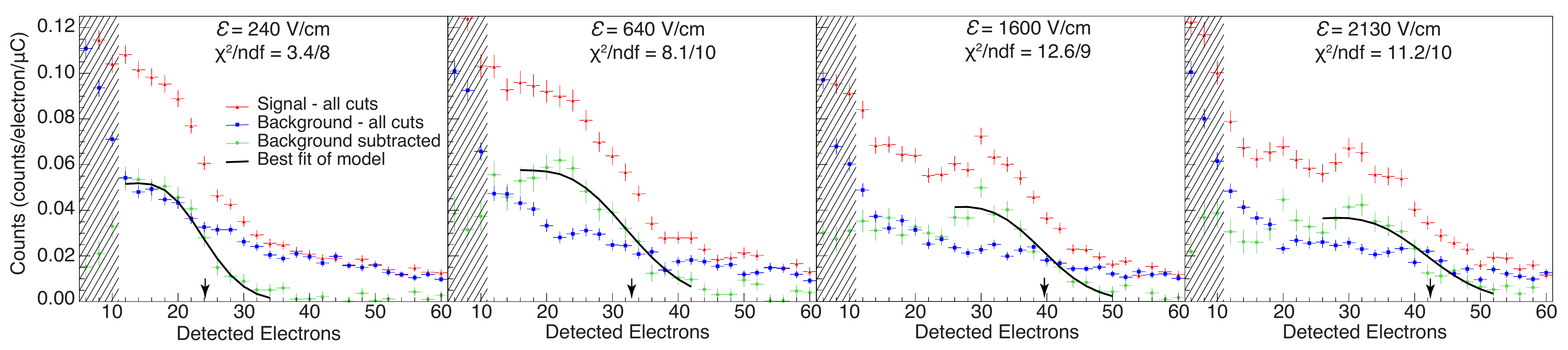}
\caption{\small{
Ionization spectrum from neutron scattering at $\mathcal{E}$=240, 640, 1600 and 2130 V/cm.  
Data quality cuts are described in the text.  
The best fit model is plotted in the fit region only.  
The best fit 6.7 keV endpoint location is indicated with an arrow.  
The efficiency of the fiducial $(x,y)$ cut begins to vary below 11 electrons (shaded region).
}}
\label{fig:Spectra}
\end{figure*}

Particle interactions in the liquid argon can produce primary scintillation and ionization. 
The detector was optimized for detection of the proportional scintillation resulting from extracting the electrons into the gas-phase argon, and accelerating them across a 1.8-cm gap. 
The detector has been shown to be sensitive to the signal resulting from a single electron \cite{Sangiorgio201369}. 
The applied electric field used to create the proportional scintillation was a constant 9.8~kV/cm for these measurements. 
The applied electric field ($\mathcal{E}$) across the liquid argon target, oriented in the $z$ direction, was varied from 240~V/cm to 2130~V/cm, in order to explore the effect on the available signal.  
Electrostatic simulations show a 6\% nonuniformity in the applied electric field within the LAr target volume, arising from the field cage spacing.  

The data acquisition was triggered by fourfold coincidence of the four phototubes, in a $10-\mu$s window. 
The trigger efficiency was consistent with unity for signals larger than 8 ionization electrons. 
New triggers were vetoed for 3 ms following very large [$>\approx$10,000 photoelectrons (PEs)] events, to exclude phototube saturation effects from the data.  

Research-grade argon was condensed into the detector through a getter to remove electronegative impurities, and a free electron lifetime of $>$ 300 $\mu$s was verified throughout the experiment as in Ref. \cite{Sangiorgio201369}.
The maximum electron drift time across the target region varied from 32 $\mu$s at 240 V/cm to 14 $\mu$s at 2130 V/cm applied electric field, leading to a mean electron loss of 5\%.  
During operation, the argon vapor pressure was maintained at 1.08 bar with 1\% stability, and the liquid temperature at approximately $88$~K (corresponding to a liquid density of 1.39~g/cm$^3$).  

$Q_y$ was measured in an end-point-type experiment.  
Monoenergetic neutrons with well-defined energy ($E_n$) interact within the liquid argon target producing nuclear recoils.  
For \textit{S}-wave scatter, expected for this experiment, nuclear recoils are populated uniformly in energy from zero to $E_{max}=4E_{n}m_{Ar}m_{n}/(m_{Ar}+m_{n})^{2}=6.7$~keV for $E_n=70$~keV scattering on \iso{40}{Ar}, the most abundant argon isotope.  
The end point in the observed ionization spectrum is then attributed to $E_{max}$.  

Quasimonoenergetic neutrons were produced with a collimated near-threshold \iso{7}{Li}(\textit{p,n})\iso{7}{Be} source, and filtered with a 7-cm length of high-purity iron as described in \cite{nsource}. 
The iron neutron filter has transmission notches at 24, 70, and 82~keV. 
The 70-keV notch was selected to target the low-energy side of the elastic neutron scattering resonance centered at 77~keV in \iso{40}{Ar} thus producing a large interaction rate while limiting the probability of multiple scatter.  

The proton beam energy was 1.932 MeV for all measurements and calibrated before and during data taking. 
Beam current was nominally 700 nA throughout data taking.  
The collimation aperture subtends $\pm1\degree$.  
The iron-filled collimator was oriented at $45 \pm0.5\degree$ with respect to the proton beam when collecting signal data.  
Representative background data were acquired at an angle of $55 \pm0.5\degree$, in which case 70~keV neutron production is kinematically forbidden, but all other beam-related backgrounds, including the 24 keV component of the neutron beam and beam-related gammas, are present.

Following the collection of neutron scatter data, a small amount of argon gas ($<$0.5g) containing 3$\pm$0.5 kBq of \iso{37}{Ar} was injected into the detector and allowed to diffuse for one hour.  
Calibration data as described in Ref. \cite{Sangiorgio201369} wer then acquired in the same four electric drift-field configurations.  

\textit{Analysis}.--
Triggered proportional scintillation (ionization channel) events identified by the analysis were subjected to a series of quality cuts.  
The cuts included the selection of {\it(a)} isolated events, defined as having $<$2 PEs in the 50-$\mu$s pre-trigger and $<$10 PEs following the event, and {\it(b)} the rejection of primary scintillation from peripheral background events, which have a characteristic fast rise and $1.6~\mu$s decay time.  
Additional cuts include the rejection of {\it(c)} events near the $(x,y)$ edge of the active region using the same algorithm described in Ref. \cite{Sangiorgio201369} and {\it(d)} pileup events, e.g., axially ($z$) separated multiple scatters, by accepting events with 95\% of signal arriving in $<20~\mu$s.  Cut {\it(c)} also strongly limits the acceptance of pile up and multiple scatters separated in $(x,y)$. 
The energy dependence of this suite of selection criteria was found to vary by $<$5\% for events with $>11$ detected electrons. 
The nuclear recoil endpoint ``shoulder'' is clearly visible before background subtraction (Fig.~\ref{fig:Spectra}).  

Fluctuations in the phototube response were less than 2\% over individual data sets. 
Single PE calibrations were performed for each data set using isolated single PE from the tail of proportional scintillation events.

The transformation of neutron scattering data from measured PE to detected electrons required a single-electron calibration from previous data because single electrons were not observed in sufficiently high rates during this experiment.  
Previous measurements with this detector found $7.8\pm0.1$~PEs per detected electron ($\mbox{PE}/e^-$) with a systematic uncertainty of 10\% due to the difficulty in localizing the $(x,y)$ coordinates of the single-electron signals.  
A value of $10.4\pm0.2$~$\mbox{PE}/e^-$ was used in the present analysis. 
The $33\%$ increase in light yield resulted from a larger electric field and physical gap in the proportional scintillation region, and was obtained using the 2.82-keV peak from \iso{37}{Ar} \textit{K}-capture (2.82 keV released in x rays/Auger electrons \cite{Barsanov}) acquired across a range of electric field configurations. 
The statistical and systematic uncertainties of this calibration were 2\% and 10\%, respectively.

Backgrounds during these measurements were dominantly beam related--namely, 24 keV neutrons that transit the iron filter, gammas from \iso{7}{Li}(\textit{p,p'})\iso{7}{Li} within the lithium target, and neutron-capture gammas--and were proportional to the proton current on the target. 
Data were normalized by the integrated proton current and corrected for the live time fraction of the data acquisition system.  
The normalized spectra were then subtracted as shown in Fig.~\ref{fig:Spectra}.  

\begin{figure}[t]
\includegraphics[angle=0,width=0.48\textwidth]{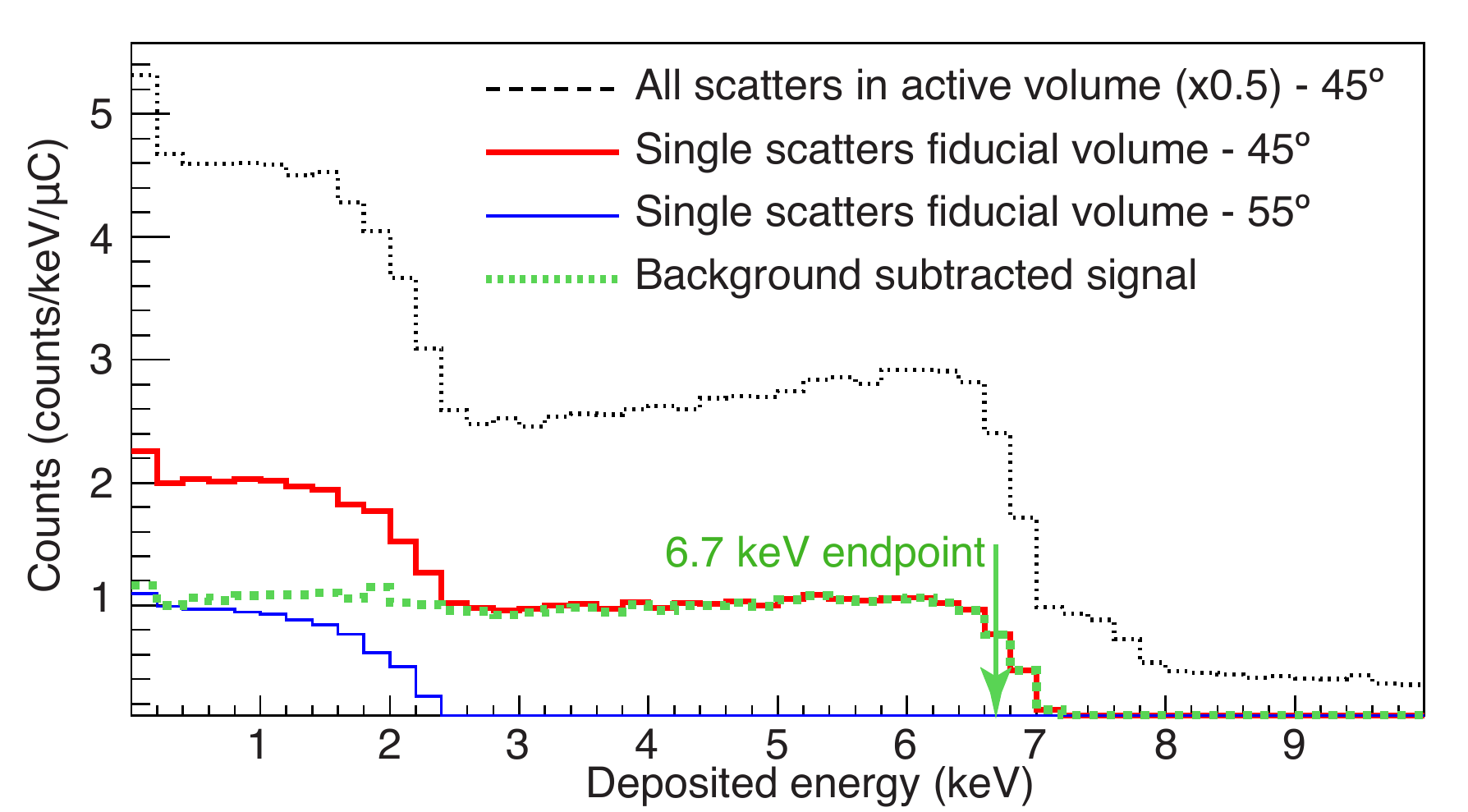}
\caption{\small{
Simulated neutron energy deposition spectra (from MCNP-Polimi) in the active detector region (dashed), and single scatter energy deposition in the fiducial detector region when the collimator is oriented at 45$\degree$ (solid thick).  
In this configuration both 70 and 24 keV neutron beams are impinging on the detector.  
At 55$\degree$ (solid thin) only 24 keV neutrons contribute.  
Background subtracted fiducial data (dotted) illustrates the experimental design to isolate the contribution of 70 keV neutrons.
}}
\label{fig:MCNP}
\end{figure}

\begin{figure}[b]
\includegraphics[angle=0,width=0.48\textwidth]{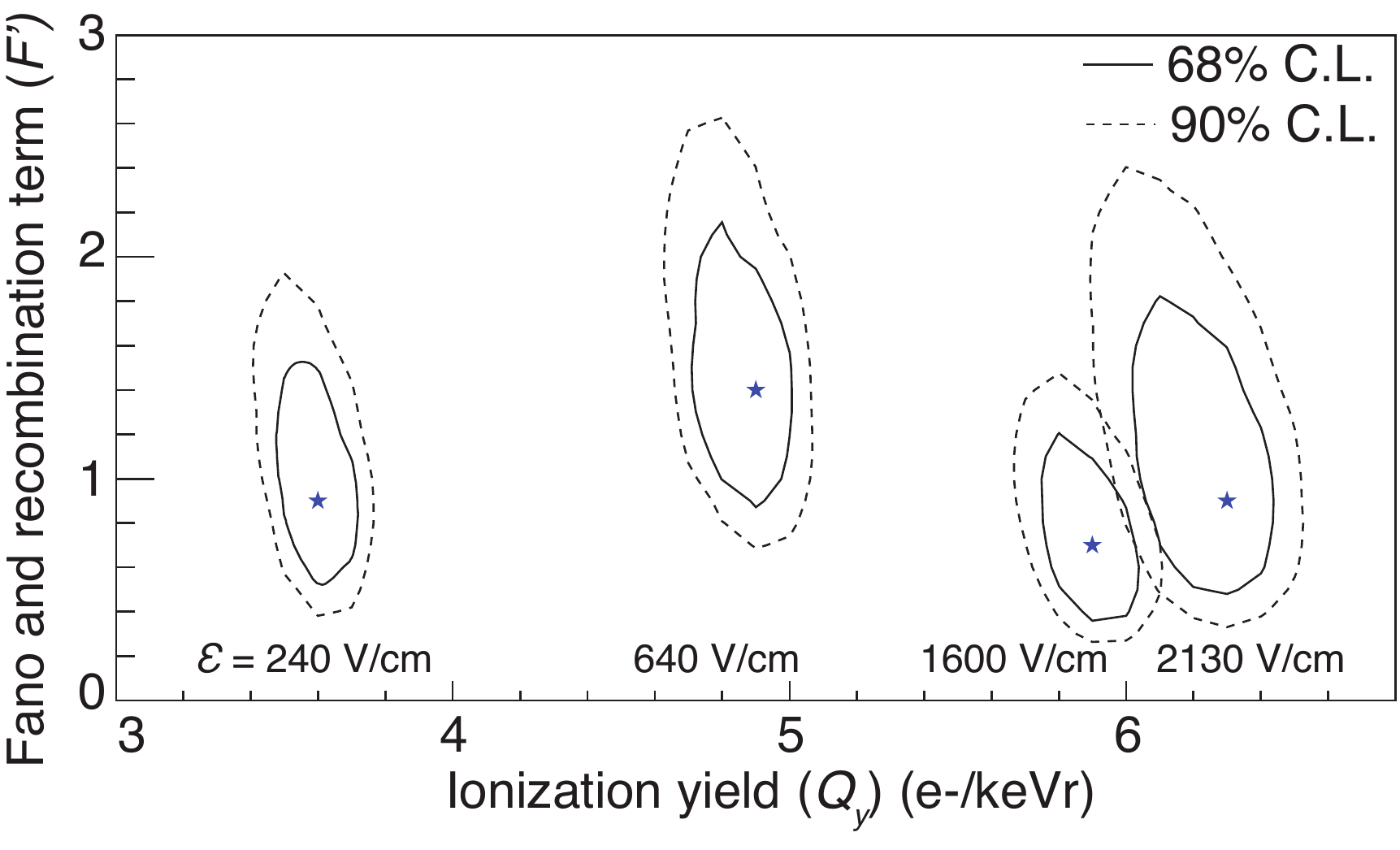}
\caption{\small{
68\% and 90\% confidence level contours are shown in the $Q_y - F'$ plane.  
Stars indicate the minimum $\chi^{2}$.
}}
\label{fig:Endpoint}
\end{figure}

A detailed MCNP-PoliMi \cite{Pozzi2003550} simulation, using the \small{ENDF}/\small{B}-VII.1 library, was performed to model the expected single-scatter spectra in both the signal and background detector configurations, as shown in Fig.~\ref{fig:MCNP}.  
For comparison with data as shown in Fig.~\ref{fig:Spectra}, the simulated spectra were first converted from recoil energy to a number of electrons via a constant ionization yield ($Q_y$). 
Then a resolution term was applied, defined as $\sigma(n_e)=\sqrt{n_e(F'+\sigma_{e}^{2})}$, where $n_e$ is the number of detected electrons and $\sigma_{e}=0.37$ is the measured single-electron resolution. 
The term $F'\equiv F+R$ accounts for the Fano factor ($F$) and recombination fluctuations ($R$). The third free parameter in the fit was the rate normalization.

A $\chi^{2}$ comparison between the simulation and the background-subtracted spectrum was made using a parametric scan across the free parameters ($Q_y$, $F'$, and rate normalization), resulting in the confidence level contours shown in Fig. \ref{fig:Endpoint}.  
The region of interest for each drift field was selected to focus on the location of the end-point shoulder.
The statistical uncertainty of the best-fit $Q_y$ value was defined by the extent of the 68\% confidence level contours.  

\begin{table}[t]
\caption{\small{Uncertainties in ionization yield ($Q_y$) endpoint analysis.}}
\begin{center}
\begin{tabular}{lcc}
\hline \hline
\emph{Component}  &  \emph{Statistical (\%)}   &  \emph{Systematic (\%)}  \\
\hline \hline
Single electron peak & 2--10 & 10 \\
Single electron calibration & 2 & 10 \\
\hline 
$\chi^{2}$ analysis & 3--5 & - \\
Input spectrum & - & 5 \\ 
Background subtraction & - & 1--3 \\
Slope of $Q_y$ in model ~~ 240 V/cm & - & $^{+5}_{-25}$ \\ [0.4ex]
~~~~~~~~~~~~~~~~~\texttt{"}~~~~~~~~~~~~~~~~~~ 640 V/cm& - & $^{+2}_{-18}$ \\ [0.4ex]
~~~~~~~~~~~~~~~~~\texttt{"}~~~~~~~~~~~~~~~~~1600 V/cm& - & $^{+0}_{-19}$ \\ [0.4ex]
~~~~~~~~~~~~~~~~~\texttt{"}~~~~~~~~~~~~~~~~~2130 V/cm & - & $^{+0}_{-21}$  \\ [0.4ex]
\hline
Liquid argon purity & - & 5 \\
\hline
Drift field ($\mathcal{E}$) & - & 6 \\
\hline 
\end{tabular}
\end{center}
\label{tab:Uncertainty}
\end{table}

We emphasize that this analysis was focused solely on extracting the ionization yield at the end point and makes no attempt to extract information about ionization yields below 6.7 keV.  
This is because at energies below the end point, it is not possible to uniquely resolve the degeneracy between the free parameters in the model. 
The most robust method of accessing information about $Q_y$ at smaller recoil energies is to decrease the end-point energy \cite{nsource}.
To estimate the systematic uncertainty associated with the assumption that $Q_y$ is constant with recoil energy, we repeated the analysis for each data set with the linear slope of the ionization yield as an additional free parameter. 
For all but the smallest value of $\mathcal{E}$, the best fit was obtained for a slope of about $-0.8~Q_{y}/\mbox{keV}$ and a slightly lower end-point $Q_y$. 
This is quoted as a systematic uncertainty for each drift field in Table ~\ref{tab:Uncertainty}.
Additionally, we repeated the analysis using a simple step function for the input nuclear recoil spectrum, to approximate the ideal \textit{S}-wave recoil spectrum from monoenergetic 70 keV neutrons (this is not shown in Fig.~\ref{fig:MCNP}).  
This provided a conservative approximation of the uncertainty due to underlying uncertainties in the differential nuclear cross-section data, used in the MCNP-Polimi simulation. 
The systematic uncertainty associated with subtraction of background data was assessed using an exponential fit to background data ($>11$ electrons).  
Using the best-fit exponential for subtraction yielded the same best-fit $Q_y$.  
Varying the exponential constant $\pm15\%$ resulted in a $\pm$1--3\% shift in best-fit $Q_y$.  

Table \ref{tab:Uncertainty} summarizes the statistical and systematic uncertainties present in the ionization yield results. 
The statistical uncertainty of the best-fit mean is quoted. 
Asymmetric uncertainties were attributed to several of the listed parameters as a result of their underlying nature.  
Uncertainties were added in quadrature when combined.

\begin{figure}[t]
\includegraphics[angle=0,width=0.48\textwidth]{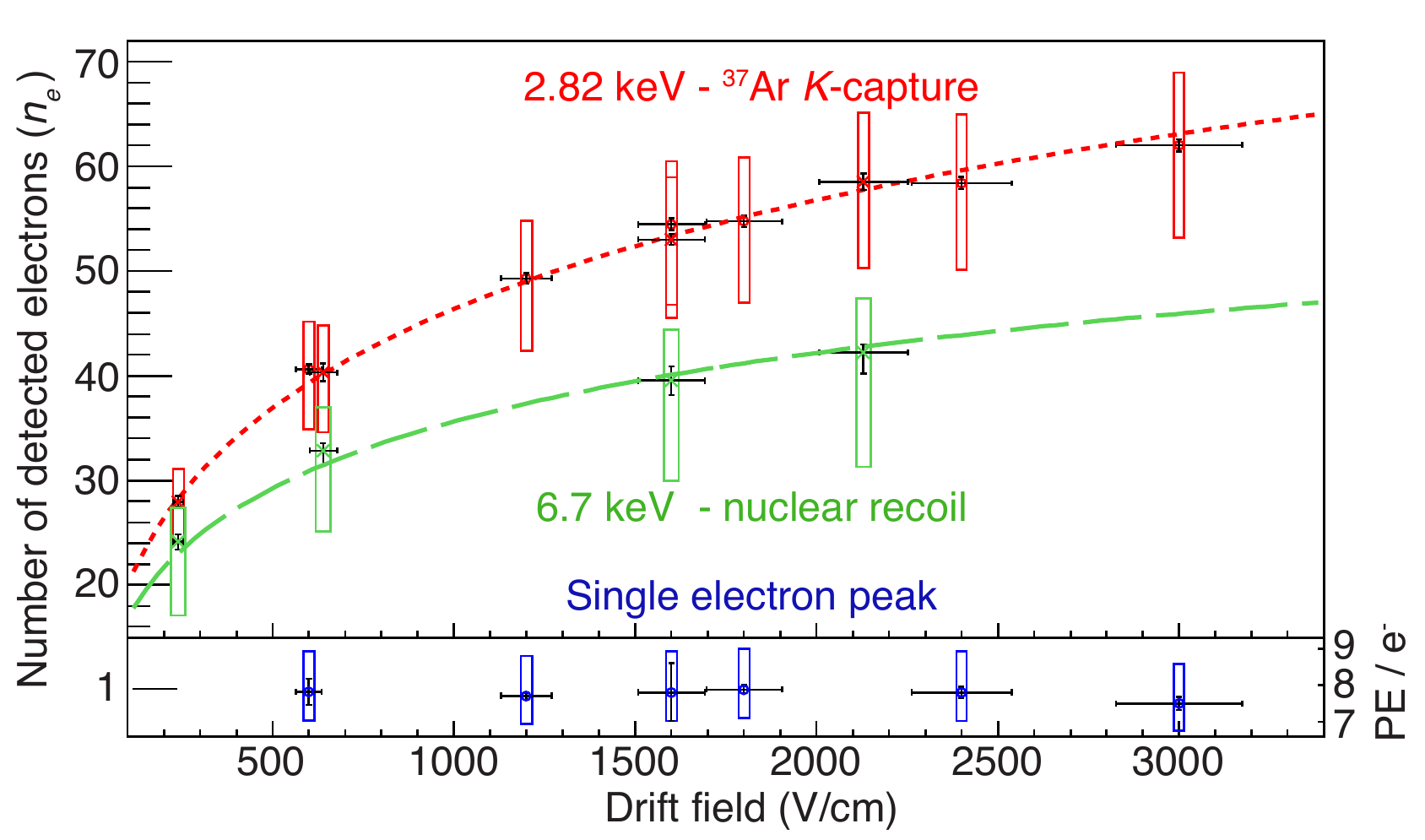}
\caption{\small{{
\it(upper)} The number of observed electrons from 2.82 keV \iso{37}{Ar} K-capture events and 6.7 keV nuclear recoils as a function of $\mathcal{E}$, with systematic (boxes) and statistical (bars) uncertainty on the mean. 
Curves are the best fit obtained from Eq. \ref{eq:TImodel}.  
{\it(lower)} The single electron peaks used (in conjunction with 2.82 keV data) to infer the single electron calibration for endpoint analysis. 
}}
\label{fig:YieldComparison}
\end{figure}

\textit{Results and Discussion}.--
The number of electrons detected from 6.7-keV nuclear recoils as a function of applied electric drift field is shown in Fig.~\ref{fig:YieldComparison} and the ionization yield with uncertainties is listed in Table~\ref{tab:Results}.  
The strong dependence on the electric drift field is in reasonable agreement with recent observations in the scintillation channel \cite{Alexander}, consistent with the expected anticorrelation of scintillation and ionization. 
The different recoil energies and the lack of absolute scintillation yields in Ref. \cite{Alexander} prevent a quantifiable comparison.  

This field dependence is understood to be a suppression of ion-electron recombination along the ionization track and is extensively discussed in Ref. \cite{Chepel}.  
In order to fit our data we consider an empirical modification \cite{Dahl, NEST} of the Thomas-Imel box model \cite{PhysRevA.36.614},
\begin{equation}
\label{eq:TImodel}
n_e=\frac{N_i}{\xi }\ln (1+\xi ),~~\xi=\frac{N_iC}{\mathcal{E}^{b}}.
\end{equation}
$N_{i}$ is the number of initial ion-electron pairs produced, $n_e$ is the number of electrons that escape recombination, $\mathcal{E}$ is the applied electric field, and $b$ and $C$ are constants.  
The electric drift-field dependence is modified from the original model to have a power-law dependence, $\xi\propto \mathcal{E}^{-b}$.  The number of initial ion-electron pairs may be written as 
\begin{equation}
\label{eq:ni}
N_i=\frac{f E}{\epsilon \left(1+N_{ex}/N_{i}\right)},
\end{equation}
where $E$ is the amount of energy deposited in the track, $f$ is the fraction of energy lost through ionization and atomic excitation (unity for electronic recoils) often termed a quench factor, $\epsilon=19.5$~eV is the average energy required to produce a quantum (excitation or ionization) in liquid argon \cite{Doke}, and $N_{ex}$ is the number of initial excitations.  
The ratio $N_{ex}/N_{i}=0.21$ was measured for electronic recoils in liquid argon \cite{Kubota}.  
The model has only two free parameters ($C,b$) when describing electron recoils.  
Using the 2.82-keV \iso{37}{Ar} \textit{K}-capture calibration data a best fit (Fig.~\ref{fig:YieldComparison}) yields $C=2.37$ and $b=0.61$ when $\mathcal{E}$ is expressed in V/cm. 

Using these values for $b$ and $C$, the number of initial ion-electron pairs ($N_i$) is left as a single free parameter when applied to nuclear recoil data.  
Fitting to the data (Fig.~\ref{fig:YieldComparison}) we observe good agreement and find $N_i=72\pm2$, assuming this model remains valid at high (saturating) field values. 
The fact that recombination in liquid argon can be described by the same phenomenological model for few-keV electron and nuclear recoils suggests a similarity in the spatial distribution of electrons and ions for these different energy-deposition mechanisms.  

Using Eq.~(\ref{eq:ni}) and the calculations of Lindhard \textit{et al.} \cite{lindhard1963integral} for the partitioning of nuclear recoil energy ($f=0.25$) results in $N_{ex}/N_{i}=0.19$, which is surprisingly similar to the value for electron recoils.  
Alternatively, if $N_{ex}/N_{i}\sim$1 (as measured for nuclear recoils in liquid xenon \cite{SandD}) then one would find $f=0.42$. 
If confirmed this would suggest a promising sensitivity of liquid argon at low energies.  
Simultaneous measurements of scintillation and ionization are needed to unambiguously determine $f$ and $N_{ex}/N_{i}$.

\begin{table}[t]
\caption{\small{Measured ionization yields with uncertainties.}}
\begin{center}
\begin{tabular}{cccc}
\hline \hline
$\mathcal{E}$ (V/cm)   & $Q_{y}$ ($e^{-}$/keV) & Statistical & Systematic \\
\hline \\[-2.3ex]
240 & 3.6 &    $^{+0.1}_{-0.1}$ & $^{+0.5}_{-1.1}$ \\    [0.4ex]
640 & 4.9 &    $^{+0.1}_{-0.2}$  & $^{+0.6}_{-1.2}$  \\  [0.4ex]
1600 & 5.9 &  $^{+0.2}_{-0.2}$  & $^{+0.7}_{-1.4}$  \\  [0.4ex]
2130 & 6.3 &  $^{+0.1}_{-0.3}$  & $^{+0.8}_{-1.6}$  \\  [0.4ex]
\hline 
\end{tabular}
\end{center}
\label{tab:Results}
\end{table}

We are not aware of any measurements or theoretical expectations for either the Fano factor or recombination fluctuations for nuclear recoils in liquid argon. 
With the simple assumption that recombination statistics are binomial, the probability for an electron to escape recombination is $p=n_e/N_i$, and so $R = 1-n_e/N_i$. 
From this, it would follow that the Fano factor is given by $F=F'+n_{e}/N_{i}-1$.  
Taking the range of $p$ from Fig. \ref{fig:YieldComparison} it is clear that $F$ is smaller than $F'$ by a factor which ranges from 0.65 at  $\mathcal{E}=240$~V/cm to 0.42 at $\mathcal{E}=2130$~V/cm. 
This is consistent with $F\approx0.5$, with a fairly large uncertainty as shown in Fig. \ref{fig:Endpoint}.

In this Letter we have presented the first nuclear recoil ionization yield measurement and the first measurement of sub-10-keV nuclear recoils in liquid argon using an end-point-type measurement.  
This demonstration suggests that end-point measurements with filtered neutron sources \cite{nsource,Barbeau,RussianFilter} are suitable for a comprehensive study of both scintillation and ionization yields of low-energy nuclear recoils in liquid argon, and could also probe $<$ 4 keV in liquid xenon.  
The results of such a study would clarify the threshold and calibration of liquid noble-based dark matter detectors and CENNS searches.  
The measurements presented in this Letter demonstrate a large ionization yield for nuclear recoils at energies below current thresholds of liquid argon dark matter searches, suggesting the ionization channel as a means for exploring light-mass dark matter in existing and future liquid argon detectors.  

\begin{acknowledgments}
We would like to thank G. Bench and T. Brown for assistance and support throughout the beam measurements and J. Coleman and K. Mavrokoridis for previous detector contributions.
We would like to acknowledge the Lawrence Scholars Program and the Department of Homeland Security for funding T.H.J.'s research. 
A portion of M.F.'s research was performed under the Nuclear Forensics Graduate Fellowship Program, which is sponsored by the U.S. Department of Homeland Security, Domestic Nuclear Detection Office, and the U.S. Department of Defense, Defense Threat Reduction Agency. 
We gratefully acknowledge the LDRD program (LDRD 13-FS-005) at LLNL.  
This work was performed under the auspices of the U.S. Department of Energy by Lawrence Livermore National Laboratory under contract DE-AC52-07NA27344. 
LLNL-JRNL-646478
\end{acknowledgments}

\newpage
\bibliographystyle{apsrev4-1}
\bibliography{EndpointBib}

\end{document}